\documentclass[10pt,twocolumn,showpacs,amsmath,amssymb,osajnl,floatfix]{revtex4-1}
\usepackage{epsfig}
\usepackage{amsmath}
\usepackage{amssymb}
\usepackage{bm}
\usepackage{graphicx}
\usepackage{color}
\usepackage{graphicx,rotating}
\begin{document}
\author{Jian-Xing Li}\email{Jian-Xing.Li@mpi-hd.mpg.de}
\author{Karen Z. Hatsagortsyan}\email{k.hatsagortsyan@mpi-hd.mpg.de}
\author{Benjamin J. Galow}
\author{Christoph H. Keitel}\email{keitel@mpi-hd.mpg.de}
\affiliation{Max-Planck-Institut f\"{u}r Kernphysik, Saupfercheckweg 1,
69117 Heidelberg, Germany}

\bibliographystyle{apsrev4-1}

\title{Attosecond gamma-ray pulses via nonlinear Compton scattering\\
in the radiation dominated regime }

\date{\today}

\begin{abstract}

 The feasibility of generation of bright ultrashort gamma-ray pulses is demonstrated in the interaction of a relativistic electron bunch with a counterpropagating tightly-focused superstrong laser beam in the radiation dominated regime.  The Compton scattering spectra of gamma-radiation are investigated using a semiclassical description for the electron dynamics in the laser field and a quantum electrodynamical description for the photon emission. We demonstrate the feasibility of ultrashort gamma-ray bursts of hundreds of attoseconds and of dozens of megaelectronvolt photon energies  in the near-backwards direction of the initial electron motion.  The tightly focused laser field structure and radiation reaction are shown to be responsible for such short gamma-ray bursts  which are independent of the durations of the electron bunch and of the laser pulse. The results are measurable with the laser technology available in a near-future.


\end{abstract}

\maketitle

Short after the invention of the laser, it was realized that Compton scattering \cite{Compton_1923} of the laser radiation by a relativistic electron beam can be a bright source of x- and gamma-rays \cite{Milburn_1963,Arutyunian_1963}. Later  proof-of-principle experiments \cite{Ting_1995,Schoenlein_1996} showed generation of picosecond hard x-rays  using  electron beams from a linear accelerator. With the appearance of the laser wakefield acceleration technique for electrons \cite{Esarey_2009}, an all-optical setup for  Thomson/Compton radiation sources from a few hundreds of keV up to
8-9 MeV  photon energies, with a shorter duration of about 50 fs, has been demonstrated \cite{Brown_2004,Schwoerer_2006L,Gibson_2010,Phuoc_2012,Jochmann_2013,Powers_2014}. 
These experiments are based on linear Thomson/Compton scattering, which produces  narrowband gamma-radiation sources for nuclear resonance fluorescence \cite{Albert_2010l}. 
Recently, a successful effort was accomplished towards Thomson/Compton scattering in the nonlinear regime \cite{Sarri_2014}. 
In superstrong laser fields Thomson/Compton scattering acquires nonlinear characteristics due to multiple laser photon absorption \cite{Goldman_1964,Brown_1964}.
Moreover, radiation reaction can enter into play in these extreme conditions  \cite{RMP_2012}, which has attracted  considerable attention recently \cite{Koga_2005,DiPiazza_2009,DiPiazza_2010,Sokolov_2010, Sokolov_2010b,Thomas_2012,Green_2014,Blackburn_2014,Jian-Xing_2014} due to present and next generation petawatt laser systems \cite{ELI,HiPER}.  

In linear Thomson/Compton scattering the duration of the emitted gamma-radiation pulse is  determined by the shortest of either the laser or electron beam duration. In an all-optical setup the electron bunch length is of the order of the laser pulse length and the created gamma-rays are of duration of several tens of femtoseconds. Are shorter pulses of gamma-rays necessary? 
Generally, short laser pulses are required for the time-resolved monitoring and control of fast-evolving processes with the pump-probe technique. The state-of-the-art time-resolution has achieved the attosecond scale by using extreme ultraviolet radiation, which allows to track the dynamics of an electronic wave packet in an atom \cite{Krausz_2009}. The required frequencies of the short pulses depend on the characteristic energies of the processes under investigation. The molecular dynamics and chemical reactions  can be controlled with a few electronvolt excitations driven by an infrared laser field, and the inner-shell electron dynamics by photons with a few 100 eV up to several keV energies. The next challenge is to time-resolve the intra-nuclear dynamics \cite{Ledingham_2003,Palffy_2015}.  It is known \cite{Povh_book} that typical energies of nuclear single-particle transitions are of the order of 1-10 MeV with typical decay lifetimes of the levels around $10^{-9}-10^{-15}$ s. The energies of the collective nuclear excitations range from several dozens of keV up to 30 MeV. The disintegration time of compound nuclei during nuclear reactions ranges from $10^{-19}-10^{-16}$ s. This sets the scale for the required photon energy and pulse duration. 
There is a wealth of nuclear phenomena for which the investigation of the time resolved dynamics requires short photon pulses, such as, resonance fluorescence  (1 fs timescale), resonance internal conversion (1 as timescale) and compound nuclei evolution (zeptosecond timescale).

\begin{figure}[b]
\includegraphics[width=8cm]{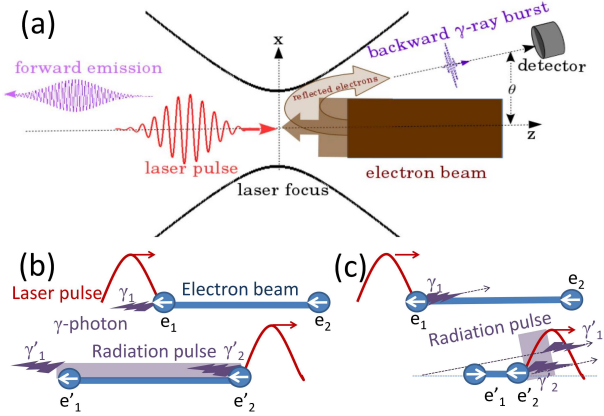}
\caption{(Color online)  The schematic representation for the generation of ultrashort gamma-ray bursts: (a) The electron  beam  counterpropagates with the laser pulse and a small slice of the electron beam looses enough energy due to radiation reaction to be reflected and emits ultrashort gamma-rays when leaving the laser focal region. (b) and (c)  show how the radiation pulse arises, when the electron beam counterpropagates (b) [or is reflected and copropagates (c)] with the laser pulse. The front part of the gamma radiation ``$\gamma_1$" is built up when the laser pulse reaches the front of the electron beam ``$\rm e_1$".  The gamma radiation pulse ends ``$\gamma'_2$" when the laser pulse overtakes the end of the electron beam ``$\rm e'_2$" (``$\gamma'_1$" and ``$\rm e'_1$" are the positions of the front of the gamma-ray and the electron beam  at this moment, respectively). The length of the gamma-ray pulse ($\gamma'_1,\gamma'_2$) is much shorter in the case of (c), because ``$\gamma_1$" and ``$\rm e_2$" counterpropagate. See more in the text and in \cite{Suppl_material}.  
}
\label{schematic}
\end{figure}

In this Letter, we investigate the feasibility of generating multi-MeV gamma-rays of several hundreds attoseconds duration via nonlinear Compton scattering of an intense laser pulse by a  counterpropagating  electron beam. The aim is to produce ultrashort gamma-rays even though using a much longer driving laser pulse and electron bunch. We find an interaction regime when only a small fraction of the electron beam looses sufficient energy due to radiation reaction and is reflected. The reflected electrons emit gamma-rays closer to the laser propagation direction during a short time while leaving the laser focal region, as shown in Fig.~\ref{schematic}. The radiation pulse is especially short because the propagation direction of its front part is opposite to that of the rear part of the electron beam (Fig.~\ref{schematic}(c)).
The scheme relies on the nonlinear regime of interaction, the tightly focused driving laser pulse, and the crucial effect of radiation reaction. All of these three ingredients are necessary to realise the ultrashort duration of the emitted gamma-rays determined solely by the intrinsic interaction mechanism. 

Let us determine the parameters of the applied regime. First of all,  we consider the nonlinear regime of  Compton scattering when the invariant laser field parameter is large, $\xi\gg 1$, where $\xi\equiv |e|E_0/(m\omega_0)$, $E_0$ and $\omega_0$ are the laser field and frequency, respectively, and $e$ and $m$ are the electron charge and mass, respectively  (Planck units $\hbar=c=1$ are used throughout). 
Second, the reflection of the counterpropagating electron in a  laser pulse requires a relativistic Lorentz factor $\gamma\approx\xi/2$.  In fact, the electron, initially at rest, in the laser field drifts along the laser propagation direction with the  Lorentz factor of the drift $\gamma_{ drift}=\xi/2$ \cite{Salamin_1996}. Similarly, the electron deviation angle with respect to the laser propagation direction can be estimated as $\theta\sim \xi/\gamma$, and the reflection condition corresponds to $\theta\sim 1$.
Third, the interaction has to be in the radiation-dominated regime (RDR), when the radiation losses during a laser period are comparable with the electron's initial energy, and the radiation reaction has a decisive impact on the electron dynamics. This regime is characterized by  the parameter $R \equiv\alpha \xi \chi \gtrsim 1$ \cite{RMP_2012},   which is the ratio of the radiated energy during a laser period to the electron energy. Here, $\alpha$ is the fine structure constant, and $\chi\equiv \gamma(\omega_0/m)\xi (1-\beta \cos\theta)$ the quantum strong field parameter, which determines the recoil of the electron during the photon emission with $\chi \approx \omega/m\gamma$ \cite{RMP_2012}. $\beta$ is the relativistic beta factor of the electron, $\theta$ the polar angle between the electron velocity and the laser propagation direction, and $\omega$ the emitted photon energy. The RDR is mostly accessible in the quantum regime of interaction when $\chi\gtrsim 1$ \cite{Baurichter_1997_new}. However, the RDR regime is only achievable  with extremely intense lasers $\xi\gg 1$. Thus, combining the quantum RDR conditions $R=\alpha \xi\chi \gtrsim 1$  and $\chi\approx 10^{-6}\gamma\xi\sim 1$, with the reflection condition $\gamma\sim\xi/2$, one requires $\gamma \sim \xi  \sim 10^3$. Electron beams of GeV energies ($\gamma \sim 10^3$) can be produced by the  laser-plasma acceleration technique \cite{Esarey_2009} and the laser intensities of $10^{23}$-$10^{24}$ W/cm$^2$ ($\xi \sim  10^3$) are anticipated with next generation facilities (see, e.g., \cite{ELI,HiPER}).

Our analysis in this Letter is based on Monte-Carlo simulations employing QED theory for the electron radiation and classical equations of motion for the propagation of electrons between photon emissions \cite{Suppl_material, Elkina_2011,Ridgers_2014,Green_2015}.
In superstrong laser fields $\xi\gg 1$, the coherence length of the photon emission is much smaller than the laser wavelength  and the typical size of the electron trajectory \cite{Ritus_1985} (see also \cite{Khokonov_2010}). As a result, the photon emission probability is determined by the local electron trajectory. In this case the photon emission probability in the laser field can be approximated by that of constant cross fields with the corresponding local value of the  parameter $\chi$ (this is the well-known synchrotron approximation) \cite{Baier_b_1994}, cf. \cite{Khokonov_2002a}. We employ a linearly polarized focused short  laser pulse, which is an approximate solution of Maxwell's equations with first order corrections with respect to the small parameters  $(k_0w_0)^{-1}$ and  $(\omega_0 \tau_0)^{-1}$ \cite{Jian-Xing_2014}, where $k_0$, $w_0$ and $\tau_0$ are the wave vector, the waist radius and the pulse duration of the laser beam, respectively.

\begin{figure}

\includegraphics[width=8cm]{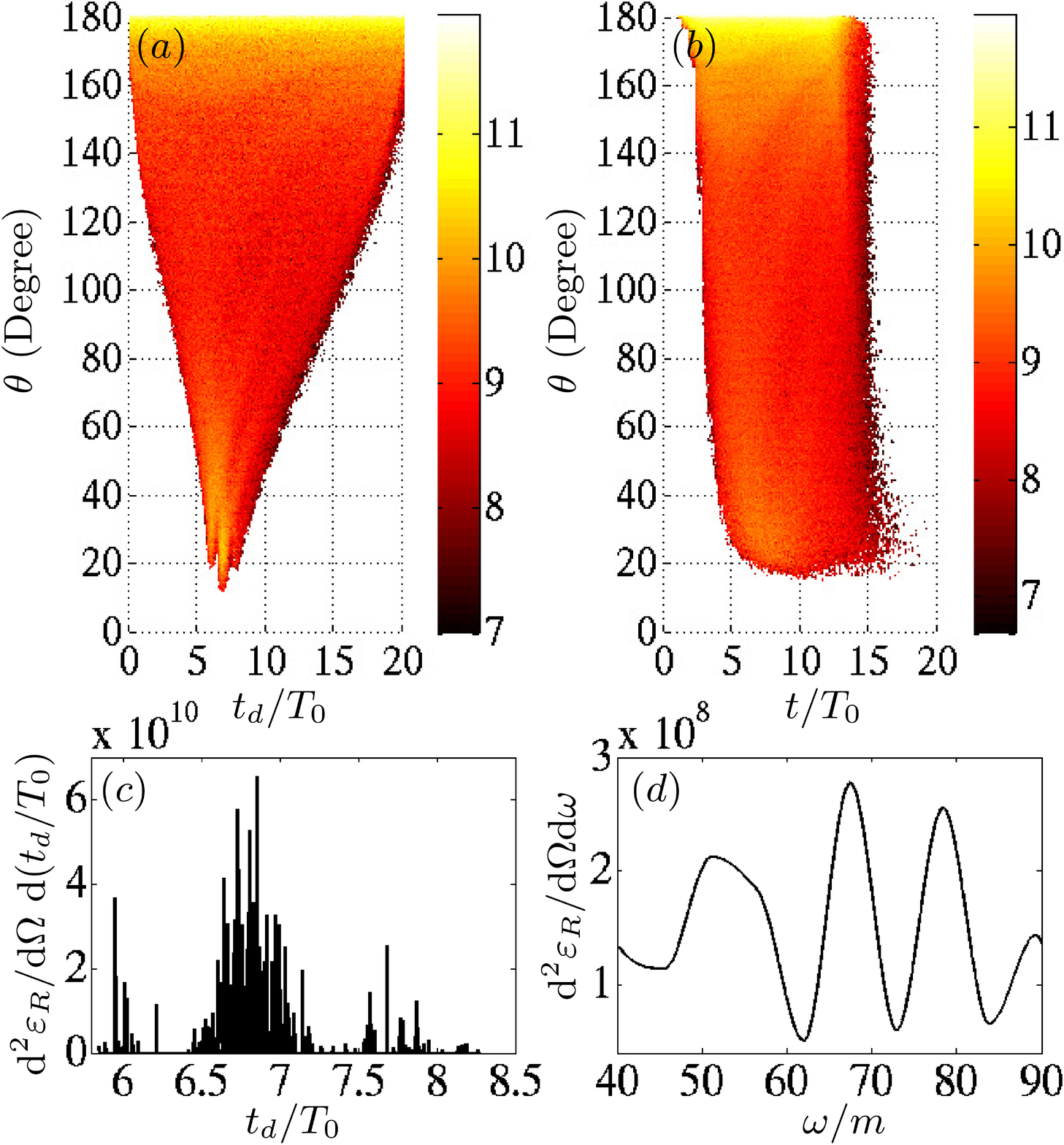}
\caption{(Color online)  The angle-resolved radiation intensity for photon energies above 1 MeV in a 4-cycle laser pulse with  carrier-envelope phase $\phi_{CEP}=0$ and azimuthal angle of emission with respect to the laser propagation direction $\phi=180^\circ$ (the spectra are similar at $\phi\approx 0$): (a) Color coded is Log$_{10}$[d$^2\varepsilon_R/$d$\Omega$d$(t_d/T_0)$] rad$^{-2}$ in the detector time $t_d$, with  the radiation energy $\varepsilon_R$ in units of the electron rest energy $m$, the laser period $T_0$, and the emission solid angle $\Omega$;  (b) Color coded is Log$_{10}$[d$^2\varepsilon_R/$d$\Omega$d$(t/T_0)$] rad$^{-2}$ with the electron emission time $t$; (c) The differential gamma-ray radiation via d$^2\varepsilon_R/$d$\Omega$d$(t_d/T_0)$ at $\theta =  20^{\circ} $ and $\Delta \theta = 0.002$ rad. 
(d) The spectral distribution d$^2 \varepsilon_R/$d$\Omega$d$\omega$ of the main pulse in (c). The laser and the electron beam parameters are given in the text.
}
\label{spectrum}
\end{figure}

The simulation results for the gamma-radiation properties above the photon energy of 1 MeV 
are shown in Fig.~\ref{spectrum}. The applied parameters are the following: the  peak intensity of the 4-cycle laser pulse is $I\approx 4.9\times 10^{23}$W/cm$^2$ ($\xi=600$), the laser wavelength $\lambda_0 = 1$ $\mu$m, and the laser beam waist size $w_0$ = 1 $\mu$m. The  initial kinetic energy of the electrons is $200$ MeV ($\gamma_0=392$, $\chi_{max}\approx0.8$). As the electron reflection condition should hold in the laser field, larger initial electron  energies  $\gamma_0>\xi/2$ are required because of radiation losses. We employ an electron bunch of length  $L_b=10\lambda_0$, and of transverse size $w_b=w_0$, with the number of electrons  $N_e=3\times 10^{8}$. The energy as well as angular spread of the  bunch are $\Delta \gamma/\gamma_0=\Delta \theta=10^{-3}$.
\begin{figure} 

\includegraphics[width=8cm]{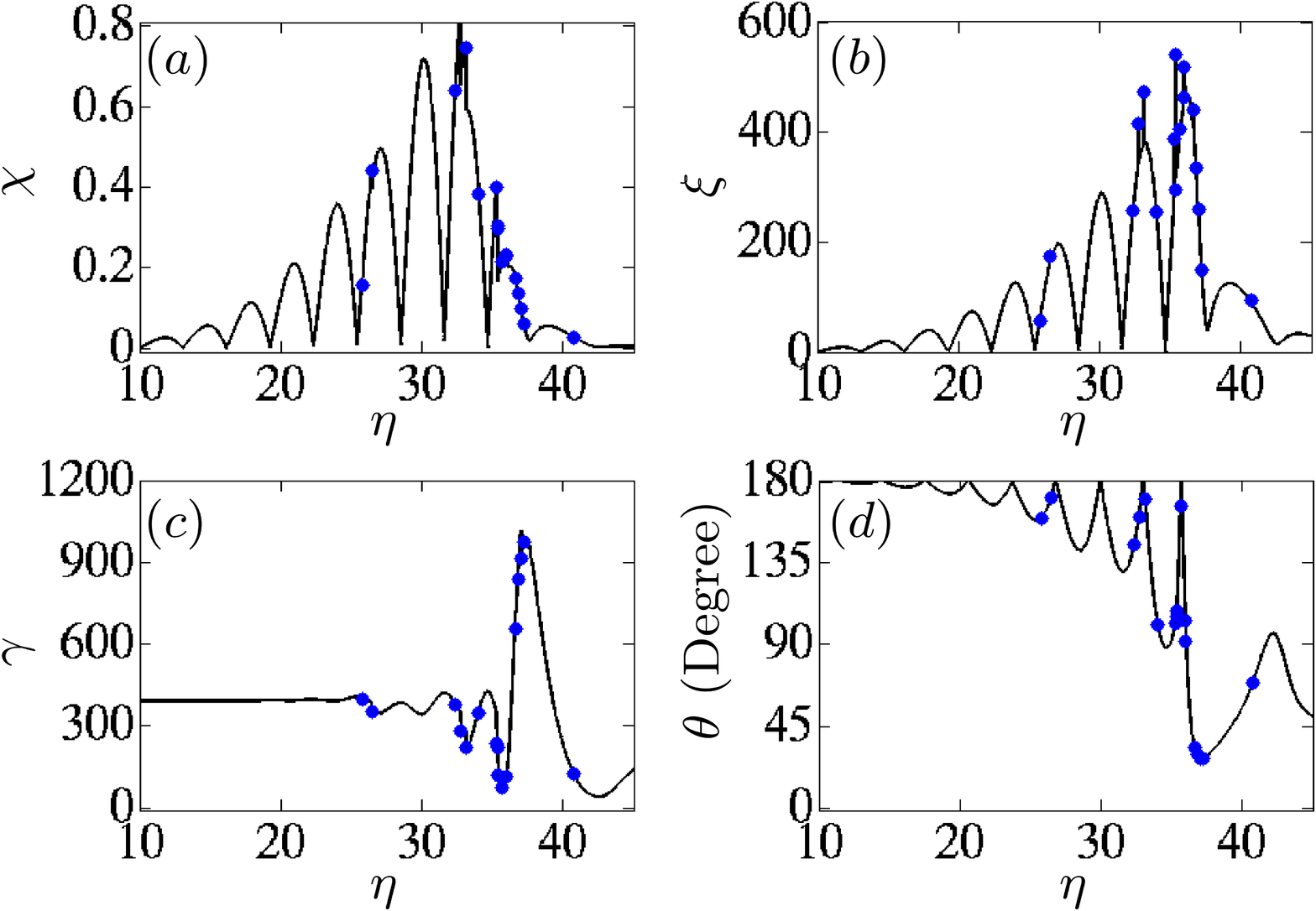}
\caption{(Color online)  Dynamics of a single exemplary electron with respect to the phase $\eta=\omega_0(t-z)$. Parameters are the same as in Fig.~\ref{spectrum}. The blue dots present the points where gamma photons are emitted. }
\label{single}
\end{figure}
 
The time-dependent angular resolved radiation intensity is shown in Fig.~\ref{spectrum}(a). The radiation sweeps from the polar angle $\theta=180^{\circ}$ down to $\theta\approx 20^{\circ}$. 
The duration of the emission decreases with decreasing $\theta$. At $\theta\approx 180^{\circ}$ it is mostly determined by the length of the electron bunch, while at small angles  it is smaller than the laser period $T_0$.  The  duration of the gamma-radiation at $\theta = 20^{\circ}$ with an aperture angle $\Delta \theta =0.002$ is illustrated in Fig.~\ref{spectrum}(c). The duration of the main gamma-pulse is about \textbf{$0.25T_0=830$} as. This is the main result of this paper: ultrashort gamma ray bursts of attosecond duration can be generated closer to the laser propagation direction, while using much longer laser and electron beams (13 fs and 33 fs, respectively, in the given example). 
The spectral distribution of the gamma ray burst is shown in Fig.~\ref{spectrum}(d), with the central frequency being $\omega \approx 67 m=34.2$ MeV. Narrower gamma-ray pulses can be detected at  smaller polar angles where, however, the mean frequency is smaller.

Let us explain the duration of the gamma radiation at different emission angles. The ultrarelativistic electrons in the bunch, which counterpropagate with the laser field, radiate initially in the direction opposite to the laser propagation [see $\eta<35$ in Fig.~\ref{single}(d)].
Significant photon emission appears when the $\chi$-parameter achieves a rather large peak value of $\chi\approx 0.6$, as shown in Fig.~\ref{single}(a). Due to the radiation loss the electron energy decreases (Fig.~\ref{single}(c)). On the other hand, at this moment the laser field is still large  (Fig.~\ref{single}(b)), yielding the electron reflection  at $\eta\approx 36$ [see the large change of $\theta$ at $\eta\approx 36$ in Fig.~\ref{single}(d)]. After the reflection the electron emits briefly closer to the laser propagation direction because it leaves the focal region with an essential decrease of the parameter $\chi$.  The emission angle  $\theta\sim \xi/\gamma\approx 20^{\circ}$, is determined by the values of $\xi$ and $\gamma$ after the reflection. This is the tilting angle of the electron trajectory  with respect  to the laser propagation direction after the reflection.

The radiation at $\theta\approx 180^{\circ}$ arises before the electrons reach the reflection point. While the duration of the radiation wave packet of a single electron is extremely short $\Delta t_d^{(1)}\sim \tau_0/4\gamma^2$, each consecutive electron in the bunch contributes into the total radiation field with corresponding time delay. Therefore, the total duration of the emission is of the order of the  electron's bunch duration $L_b/c$. An accurate estimation of the radiation pulse duration yields $\Delta t_d\sim \tau_0/ 4\gamma^2+2L_b/c(1+\beta)$, and the radiation time of the electron beam $\Delta t=(\tau_0+L_b/c)/(1+\beta)$ \cite{Suppl_material}.  Furthermore, $\Delta t_d/\Delta t\approx 1.4$ which corresponds to Fig.~\ref{spectrum}(a) and (b). The length of the emitted gamma radiation pulse is deduced by calculating the distance between the front of the gamma-pulse (which arises when the laser pulse reaches the front part of the electron beam) and the end of the gamma-pulse (which is determined by the moment  when the laser pulse reaches the end of the electron beam) \cite{Suppl_material} (see also Fig.~\ref{schematic}(b) and (c)).

\begin{figure}

\includegraphics[width=8cm]{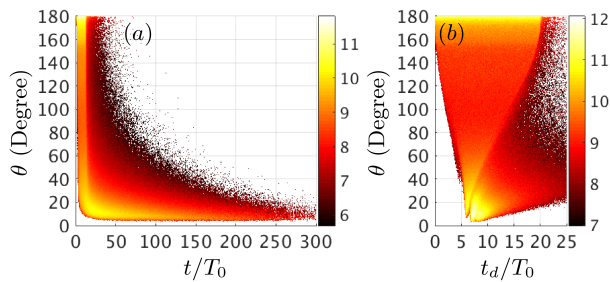}
\caption{(Color online) The angle-resolved radiation intensity in a 4-cycle plane-wave laser pulse with photons energies above 1 MeV: (a)  Log$_{10}$[d$^2\varepsilon_R/$d$\Omega$d$(t/T_0)$] rad$^{-2}$ in the electron time; (b) Log$_{10}$[d$^2\varepsilon_R/$d$\Omega$d$(t_d/T_0)$] rad$^{-2}$ in the detector time. Parameters are the same as in  Fig.~\ref{spectrum}. }
\label{plane-wave}
\end{figure}

The radiation near $\theta \approx 20^{\circ}$ arises after the reflection of the electrons. When the laser field is tightly focused, the emission time in this case again is of the order of the duration of the electron beam $\Delta t\approx L_R/c+L_b/c(1+\beta)\sim L_b/2c$, with  the Rayleigh length $L_R=\pi w_0^2/\lambda_0$ \cite{Suppl_material}.  However, due to reflection, the front of the radiation field and the rear part of the electron beam are counterpropagating, which yields a very short radiation pulse:   $\Delta t_d\sim \pi T_0(w_0/\lambda_0)^2(1-\beta\cos\theta)+L_b (1-\cos\theta)/c(1+\beta)\sim L_b (1-\cos\theta)/2c $ \cite{Suppl_material}.

Moreover, there is another reason which further decreases the emission time. During the forward radiation all electrons loose energy which facilitates the  reflection when the condition $\gamma\sim \xi/2$ is approached. However, only the front fraction of the bunch  can encounter the strongest laser field at the peak of the laser pulse within the focal region and loose enough energy to fulfil the reflection condition. The rear electrons experience a weaker laser field, because of the laser defocusing effect and can not achieve the reflection. Then, the effective length $\tilde{L}_b $ of the reflected electron bunch is shorter than  the total bunch length, for the given parameters $\tilde{L}_b\approx 10\lambda_0$ \cite{Suppl_material}. For the given parameters our estimate provides $\Delta t_d\approx 0.24T_0$  and $\Delta t\approx 4T_0$, which is in  agreement with  Fig.~\ref{spectrum}(a) and (b).

The focusing of the laser beam is absolutely essential. In a plane-wave laser pulse, as shown in Fig.~\ref{plane-wave},  the emission time of the radiation before the reflection can be seen from the area $\theta>90^{\circ}$  in Fig.~\ref{plane-wave}(a), and that after the reflection from $\theta<90^{\circ}$. In the latter case, the electron moves along the propagation direction of the laser pulse and experiences the field of the rest of the laser pulse. Then, the emission of photons takes the long time   $\Delta t\sim \tau_0/(1-\beta)\gg L_b/c$ \cite{Suppl_material}. In the frame of the detector, the duration of the radiation pulse is shortened with respect to the emission time by a factor $1-\beta \cos\theta$ to become $\Delta t_d\sim \tau_0(1-\beta\cos\theta)/(1-\beta)$ \cite{Suppl_material} [see the time range for $\theta <90^\circ$ in Fig.~\ref{plane-wave}(b) compared to Fig.~\ref{plane-wave}(a)]. However,
the shortest duration of the radiation pulse at $\theta\approx 20^{\circ}$ is still much larger than the laser period, in contrast to the case of the focused field [in the focused field $\tau_0/(1-\beta)$ is replaced by $L_R/c$ in the emission time].

Several gamma-ray bursts are observable near $\theta \approx 20^{\circ}$ in the 4-cycle laser pulse (Fig.~\ref{spectrum}(a)). Moreover, a single gamma-ray burst arises in a 2-cycle laser pulse, and a gamma-ray comb is formed by employing longer laser pulses \cite{Suppl_material}.
This is because the electron can be reflected at any wave crest due to the stochastic character of the gamma-photon emission, while there is only one single burst when stochastic effects are neglected \cite{Suppl_material}. This feature of the angle-resolved radiation intensity can serve as an indicator of stochastic effects in photon emission.

The described effect of the short gamma ray generation near the reflection condition is rather robust.
While here one example of the effect at $\xi =600$ and $\gamma_0\approx 400$ has been provided, our simulations show that the same mechanism for ultrashort gamma-ray bursts works as well when varying the laser field and the electron energy within a large range:    $\delta \xi/\xi\sim 1/2$ and $\delta \gamma/\gamma_0 \sim 1/2$  \cite{Suppl_material}.

Finally, we  estimate the total number of photons in the  gamma-ray burst of $830$ as duration for the parameters in Fig.~\ref{spectrum} to be $N_{ph}\sim  10^2$ within the emission solid angle $\Delta\Omega=1$ mrad$^2$. In spite of a small number of total photons, the photon flux (F) and the brilliance (B) are rather large due to the short duration of the pulse: $F\sim 10^{14}$ photons s$^{-1}$ $0.1 \%$BW,  and  $B\sim 3\times 10^{20}$ photons s$^{-1}$ mrad$^{-2}$ mm$^{-2}$ $0.1 \%$BW, respectively, e.g., the brilliance is 2 orders of magnitude larger than  in the recent experiment \cite{Powers_2014}.

In conclusion, we have shown that brilliant attosecond gamma-ray bursts can be produced by the combined effect of laser focusing and radiation reaction in nonlinear Compton scattering in the radiation dominated regime.

We thank Naveen Kumar and Tahir Shaaran for useful comments on this Letter. 
\bibliography{strong_fields_bibliography}

\begin{thebibliography}{42}%
\makeatletter
\providecommand \@ifxundefined [1]{%
 \@ifx{#1\undefined}
}%
\providecommand \@ifnum [1]{%
 \ifnum #1\expandafter \@firstoftwo
 \else \expandafter \@secondoftwo
 \fi
}%
\providecommand \@ifx [1]{%
 \ifx #1\expandafter \@firstoftwo
 \else \expandafter \@secondoftwo
 \fi
}%
\providecommand \natexlab [1]{#1}%
\providecommand \enquote  [1]{``#1''}%
\providecommand \bibnamefont  [1]{#1}%
\providecommand \bibfnamefont [1]{#1}%
\providecommand \citenamefont [1]{#1}%
\providecommand \href@noop [0]{\@secondoftwo}%
\providecommand \href [0]{\begingroup \@sanitize@url \@href}%
\providecommand \@href[1]{\@@startlink{#1}\@@href}%
\providecommand \@@href[1]{\endgroup#1\@@endlink}%
\providecommand \@sanitize@url [0]{\catcode `\\12\catcode `\$12\catcode
  `\&12\catcode `\#12\catcode `\^12\catcode `\_12\catcode `\%12\relax}%
\providecommand \@@startlink[1]{}%
\providecommand \@@endlink[0]{}%
\providecommand \url  [0]{\begingroup\@sanitize@url \@url }%
\providecommand \@url [1]{\endgroup\@href {#1}{\urlprefix }}%
\providecommand \urlprefix  [0]{URL }%
\providecommand \Eprint [0]{\href }%
\providecommand \doibase [0]{http://dx.doi.org/}%
\providecommand \selectlanguage [0]{\@gobble}%
\providecommand \bibinfo  [0]{\@secondoftwo}%
\providecommand \bibfield  [0]{\@secondoftwo}%
\providecommand \translation [1]{[#1]}%
\providecommand \BibitemOpen [0]{}%
\providecommand \bibitemStop [0]{}%
\providecommand \bibitemNoStop [0]{.\EOS\space}%
\providecommand \EOS [0]{\spacefactor3000\relax}%
\providecommand \BibitemShut  [1]{\csname bibitem#1\endcsname}%
\let\auto@bib@innerbib\@empty
\bibitem [{\citenamefont {Compton}(1923)}]{Compton_1923}%
  \BibitemOpen
  \bibfield  {author} {\bibinfo {author} {\bibfnamefont {A.~H.}\ \bibnamefont
  {Compton}},\ }\href@noop {} {\bibfield  {journal} {\bibinfo  {journal} {Phys.
  Rev.}\ }\textbf {\bibinfo {volume} {21}},\ \bibinfo {pages} {483} (\bibinfo
  {year} {1923})}\BibitemShut {NoStop}%
\bibitem [{\citenamefont {Milburn}(1963)}]{Milburn_1963}%
  \BibitemOpen
  \bibfield  {author} {\bibinfo {author} {\bibfnamefont {R.~H.}\ \bibnamefont
  {Milburn}},\ }\href@noop {} {\bibfield  {journal} {\bibinfo  {journal} {Phys.
  Rev. Lett.}\ }\textbf {\bibinfo {volume} {10}},\ \bibinfo {pages} {75}
  (\bibinfo {year} {1963})}\BibitemShut {NoStop}%
\bibitem [{\citenamefont {Arutyunian}\ and\ \citenamefont
  {Tumanian}(1963)}]{Arutyunian_1963}%
  \BibitemOpen
  \bibfield  {author} {\bibinfo {author} {\bibfnamefont {F.}~\bibnamefont
  {Arutyunian}}\ and\ \bibinfo {author} {\bibfnamefont {V.}~\bibnamefont
  {Tumanian}},\ }\href@noop {} {\bibfield  {journal} {\bibinfo  {journal}
  {Phys. Lett.}\ }\textbf {\bibinfo {volume} {4}},\ \bibinfo {pages} {176 }
  (\bibinfo {year} {1963})}\BibitemShut {NoStop}%
\bibitem [{\citenamefont {Ting}\ \emph {et~al.}(1995)\citenamefont {Ting},
  \citenamefont {Fischer}, \citenamefont {Fisher}, \citenamefont {Evans},
  \citenamefont {Burris}, \citenamefont {Krall}, \citenamefont {Esarey},\ and\
  \citenamefont {Sprangle}}]{Ting_1995}%
  \BibitemOpen
  \bibfield  {author} {\bibinfo {author} {\bibfnamefont {A.}~\bibnamefont
  {Ting}}, \bibinfo {author} {\bibfnamefont {R.}~\bibnamefont {Fischer}},
  \bibinfo {author} {\bibfnamefont {A.}~\bibnamefont {Fisher}}, \bibinfo
  {author} {\bibfnamefont {K.}~\bibnamefont {Evans}}, \bibinfo {author}
  {\bibfnamefont {R.}~\bibnamefont {Burris}}, \bibinfo {author} {\bibfnamefont
  {J.}~\bibnamefont {Krall}}, \bibinfo {author} {\bibfnamefont
  {E.}~\bibnamefont {Esarey}}, \ and\ \bibinfo {author} {\bibfnamefont
  {P.}~\bibnamefont {Sprangle}},\ }\href@noop {} {\bibfield  {journal}
  {\bibinfo  {journal} {J. Appl. Phys.}\ }\textbf {\bibinfo {volume} {78}}
  (\bibinfo {year} {1995})}\BibitemShut {NoStop}%
\bibitem [{\citenamefont {Schoenlein}\ \emph {et~al.}(1996)\citenamefont
  {Schoenlein}, \citenamefont {Leemans}, \citenamefont {Chin}, \citenamefont
  {Volfbeyn}, \citenamefont {Glover}, \citenamefont {Balling}, \citenamefont
  {Zolotorev}, \citenamefont {Kim}, \citenamefont {Chattopadhyay},\ and\
  \citenamefont {Shank}}]{Schoenlein_1996}%
  \BibitemOpen
  \bibfield  {author} {\bibinfo {author} {\bibfnamefont {R.~W.}\ \bibnamefont
  {Schoenlein}}, \bibinfo {author} {\bibfnamefont {W.~P.}\ \bibnamefont
  {Leemans}}, \bibinfo {author} {\bibfnamefont {A.~H.}\ \bibnamefont {Chin}},
  \bibinfo {author} {\bibfnamefont {P.}~\bibnamefont {Volfbeyn}}, \bibinfo
  {author} {\bibfnamefont {T.~E.}\ \bibnamefont {Glover}}, \bibinfo {author}
  {\bibfnamefont {P.}~\bibnamefont {Balling}}, \bibinfo {author} {\bibfnamefont
  {M.}~\bibnamefont {Zolotorev}}, \bibinfo {author} {\bibfnamefont {K.-J.}\
  \bibnamefont {Kim}}, \bibinfo {author} {\bibfnamefont {S.}~\bibnamefont
  {Chattopadhyay}}, \ and\ \bibinfo {author} {\bibfnamefont {C.~V.}\
  \bibnamefont {Shank}},\ }\href@noop {} {\bibfield  {journal} {\bibinfo
  {journal} {Science}\ }\textbf {\bibinfo {volume} {274}},\ \bibinfo {pages}
  {236} (\bibinfo {year} {1996})}\BibitemShut {NoStop}%
\bibitem [{\citenamefont {Esarey}\ \emph {et~al.}(2009)\citenamefont {Esarey},
  \citenamefont {Schroeder},\ and\ \citenamefont {Leemans}}]{Esarey_2009}%
  \BibitemOpen
  \bibfield  {author} {\bibinfo {author} {\bibfnamefont {E.}~\bibnamefont
  {Esarey}}, \bibinfo {author} {\bibfnamefont {C.~B.}\ \bibnamefont
  {Schroeder}}, \ and\ \bibinfo {author} {\bibfnamefont {W.~P.}\ \bibnamefont
  {Leemans}},\ }\href@noop {} {\bibfield  {journal} {\bibinfo  {journal} {Rev.
  Mod. Phys.}\ }\textbf {\bibinfo {volume} {81}},\ \bibinfo {pages} {1229}
  (\bibinfo {year} {2009})}\BibitemShut {NoStop}%
\bibitem [{\citenamefont {Brown}\ \emph {et~al.}(2004)\citenamefont {Brown},
  \citenamefont {Anderson}, \citenamefont {Barty}, \citenamefont {Betts},
  \citenamefont {Booth}, \citenamefont {Crane}, \citenamefont {Cross},
  \citenamefont {Fittinghoff}, \citenamefont {Gibson}, \citenamefont
  {Hartemann}, \citenamefont {Hartouni}, \citenamefont {Kuba}, \citenamefont
  {Le~Sage}, \citenamefont {Slaughter}, \citenamefont {Tremaine}, \citenamefont
  {Wootton}, \citenamefont {Springer},\ and\ \citenamefont
  {Rosenzweig}}]{Brown_2004}%
  \BibitemOpen
  \bibfield  {author} {\bibinfo {author} {\bibfnamefont {W.~J.}\ \bibnamefont
  {Brown}}, \bibinfo {author} {\bibfnamefont {S.~G.}\ \bibnamefont {Anderson}},
  \bibinfo {author} {\bibfnamefont {C.~P.~J.}\ \bibnamefont {Barty}}, \bibinfo
  {author} {\bibfnamefont {S.~M.}\ \bibnamefont {Betts}}, \bibinfo {author}
  {\bibfnamefont {R.}~\bibnamefont {Booth}}, \bibinfo {author} {\bibfnamefont
  {J.~K.}\ \bibnamefont {Crane}}, \bibinfo {author} {\bibfnamefont {R.~R.}\
  \bibnamefont {Cross}}, \bibinfo {author} {\bibfnamefont {D.~N.}\ \bibnamefont
  {Fittinghoff}}, \bibinfo {author} {\bibfnamefont {D.~J.}\ \bibnamefont
  {Gibson}}, \bibinfo {author} {\bibfnamefont {F.~V.}\ \bibnamefont
  {Hartemann}}, \bibinfo {author} {\bibfnamefont {E.~P.}\ \bibnamefont
  {Hartouni}}, \bibinfo {author} {\bibfnamefont {J.}~\bibnamefont {Kuba}},
  \bibinfo {author} {\bibfnamefont {G.~P.}\ \bibnamefont {Le~Sage}}, \bibinfo
  {author} {\bibfnamefont {D.~R.}\ \bibnamefont {Slaughter}}, \bibinfo {author}
  {\bibfnamefont {A.~M.}\ \bibnamefont {Tremaine}}, \bibinfo {author}
  {\bibfnamefont {A.~J.}\ \bibnamefont {Wootton}}, \bibinfo {author}
  {\bibfnamefont {P.~T.}\ \bibnamefont {Springer}}, \ and\ \bibinfo {author}
  {\bibfnamefont {J.~B.}\ \bibnamefont {Rosenzweig}},\ }\href@noop {}
  {\bibfield  {journal} {\bibinfo  {journal} {Phys. Rev. ST AB}\ }\textbf
  {\bibinfo {volume} {7}},\ \bibinfo {pages} {060702} (\bibinfo {year}
  {2004})}\BibitemShut {NoStop}%
\bibitem [{\citenamefont {Schwoerer}\ \emph {et~al.}(2006)\citenamefont
  {Schwoerer}, \citenamefont {Liesfeld}, \citenamefont {Schlenvoigt},
  \citenamefont {Amthor},\ and\ \citenamefont {Sauerbrey}}]{Schwoerer_2006L}%
  \BibitemOpen
  \bibfield  {author} {\bibinfo {author} {\bibfnamefont {H.}~\bibnamefont
  {Schwoerer}}, \bibinfo {author} {\bibfnamefont {B.}~\bibnamefont {Liesfeld}},
  \bibinfo {author} {\bibfnamefont {H.-P.}\ \bibnamefont {Schlenvoigt}},
  \bibinfo {author} {\bibfnamefont {K.-U.}\ \bibnamefont {Amthor}}, \ and\
  \bibinfo {author} {\bibfnamefont {R.}~\bibnamefont {Sauerbrey}},\ }\href@noop
  {} {\bibfield  {journal} {\bibinfo  {journal} {Phys. Rev. Lett.}\ }\textbf
  {\bibinfo {volume} {96}},\ \bibinfo {pages} {014802} (\bibinfo {year}
  {2006})}\BibitemShut {NoStop}%
\bibitem [{\citenamefont {Gibson}\ \emph {et~al.}(2010)\citenamefont {Gibson},
  \citenamefont {Albert}, \citenamefont {Anderson}, \citenamefont {Betts},
  \citenamefont {Messerly}, \citenamefont {Phan}, \citenamefont {Semenov},
  \citenamefont {Shverdin}, \citenamefont {Tremaine}, \citenamefont
  {Hartemann}, \citenamefont {Siders}, \citenamefont {McNabb},\ and\
  \citenamefont {Barty}}]{Gibson_2010}%
  \BibitemOpen
  \bibfield  {author} {\bibinfo {author} {\bibfnamefont {D.~J.}\ \bibnamefont
  {Gibson}}, \bibinfo {author} {\bibfnamefont {F.}~\bibnamefont {Albert}},
  \bibinfo {author} {\bibfnamefont {S.~G.}\ \bibnamefont {Anderson}}, \bibinfo
  {author} {\bibfnamefont {S.~M.}\ \bibnamefont {Betts}}, \bibinfo {author}
  {\bibfnamefont {M.~J.}\ \bibnamefont {Messerly}}, \bibinfo {author}
  {\bibfnamefont {H.~H.}\ \bibnamefont {Phan}}, \bibinfo {author}
  {\bibfnamefont {V.~A.}\ \bibnamefont {Semenov}}, \bibinfo {author}
  {\bibfnamefont {M.~Y.}\ \bibnamefont {Shverdin}}, \bibinfo {author}
  {\bibfnamefont {A.~M.}\ \bibnamefont {Tremaine}}, \bibinfo {author}
  {\bibfnamefont {F.~V.}\ \bibnamefont {Hartemann}}, \bibinfo {author}
  {\bibfnamefont {C.~W.}\ \bibnamefont {Siders}}, \bibinfo {author}
  {\bibfnamefont {D.~P.}\ \bibnamefont {McNabb}}, \ and\ \bibinfo {author}
  {\bibfnamefont {C.~P.~J.}\ \bibnamefont {Barty}},\ }\href@noop {} {\bibfield
  {journal} {\bibinfo  {journal} {Phys. Rev. ST AB}\ }\textbf {\bibinfo
  {volume} {13}},\ \bibinfo {pages} {070703} (\bibinfo {year}
  {2010})}\BibitemShut {NoStop}%
\bibitem [{\citenamefont {Phuoc}\ \emph {et~al.}(2012)\citenamefont {Phuoc},
  \citenamefont {Corde}, \citenamefont {Thaury}, \citenamefont {Malka},
  \citenamefont {Tafzi}, \citenamefont {Goddet}, \citenamefont {Shah},\ and\
  \citenamefont {Rousse}}]{Phuoc_2012}%
  \BibitemOpen
  \bibfield  {author} {\bibinfo {author} {\bibfnamefont {K.~T.}\ \bibnamefont
  {Phuoc}}, \bibinfo {author} {\bibfnamefont {S.}~\bibnamefont {Corde}},
  \bibinfo {author} {\bibfnamefont {C.}~\bibnamefont {Thaury}}, \bibinfo
  {author} {\bibfnamefont {V.}~\bibnamefont {Malka}}, \bibinfo {author}
  {\bibfnamefont {A.}~\bibnamefont {Tafzi}}, \bibinfo {author} {\bibfnamefont
  {J.~P.}\ \bibnamefont {Goddet}}, \bibinfo {author} {\bibfnamefont {R.~C.}\
  \bibnamefont {Shah}}, \ and\ \bibinfo {author} {\bibfnamefont {S.~S.~A.}\
  \bibnamefont {Rousse}},\ }\href@noop {} {\bibfield  {journal} {\bibinfo
  {journal} {Nature Phys.}\ }\textbf {\bibinfo {volume} {6}},\ \bibinfo {pages}
  {308} (\bibinfo {year} {2012})}\BibitemShut {NoStop}%
\bibitem [{\citenamefont {Jochmann}\ \emph {et~al.}(2013)\citenamefont
  {Jochmann}, \citenamefont {Irman}, \citenamefont {Bussmann}, \citenamefont
  {Couperus}, \citenamefont {Cowan}, \citenamefont {Debus}, \citenamefont
  {Kuntzsch}, \citenamefont {Ledingham}, \citenamefont {Lehnert}, \citenamefont
  {Sauerbrey}, \citenamefont {Schlenvoigt}, \citenamefont {Seipt},
  \citenamefont {St\"ohlker}, \citenamefont {Thorn}, \citenamefont {Trotsenko},
  \citenamefont {Wagner},\ and\ \citenamefont {Schramm}}]{Jochmann_2013}%
  \BibitemOpen
  \bibfield  {author} {\bibinfo {author} {\bibfnamefont {A.}~\bibnamefont
  {Jochmann}}, \bibinfo {author} {\bibfnamefont {A.}~\bibnamefont {Irman}},
  \bibinfo {author} {\bibfnamefont {M.}~\bibnamefont {Bussmann}}, \bibinfo
  {author} {\bibfnamefont {J.~P.}\ \bibnamefont {Couperus}}, \bibinfo {author}
  {\bibfnamefont {T.~E.}\ \bibnamefont {Cowan}}, \bibinfo {author}
  {\bibfnamefont {A.~D.}\ \bibnamefont {Debus}}, \bibinfo {author}
  {\bibfnamefont {M.}~\bibnamefont {Kuntzsch}}, \bibinfo {author}
  {\bibfnamefont {K.~W.~D.}\ \bibnamefont {Ledingham}}, \bibinfo {author}
  {\bibfnamefont {U.}~\bibnamefont {Lehnert}}, \bibinfo {author} {\bibfnamefont
  {R.}~\bibnamefont {Sauerbrey}}, \bibinfo {author} {\bibfnamefont {H.~P.}\
  \bibnamefont {Schlenvoigt}}, \bibinfo {author} {\bibfnamefont
  {D.}~\bibnamefont {Seipt}}, \bibinfo {author} {\bibfnamefont
  {T.}~\bibnamefont {St\"ohlker}}, \bibinfo {author} {\bibfnamefont {D.~B.}\
  \bibnamefont {Thorn}}, \bibinfo {author} {\bibfnamefont {S.}~\bibnamefont
  {Trotsenko}}, \bibinfo {author} {\bibfnamefont {A.}~\bibnamefont {Wagner}}, \
  and\ \bibinfo {author} {\bibfnamefont {U.}~\bibnamefont {Schramm}},\
  }\href@noop {} {\bibfield  {journal} {\bibinfo  {journal} {Phys. Rev. Lett.}\
  }\textbf {\bibinfo {volume} {111}},\ \bibinfo {pages} {114803} (\bibinfo
  {year} {2013})}\BibitemShut {NoStop}%
\bibitem [{\citenamefont {Powers}\ \emph {et~al.}(2014)\citenamefont {Powers},
  \citenamefont {Ghebregziabher}, \citenamefont {Golovin}, \citenamefont {Liu},
  \citenamefont {Chen}, \citenamefont {Banerjee}, \citenamefont {Zhang},\ and\
  \citenamefont {Umstadter}}]{Powers_2014}%
  \BibitemOpen
  \bibfield  {author} {\bibinfo {author} {\bibfnamefont {N.~D.}\ \bibnamefont
  {Powers}}, \bibinfo {author} {\bibfnamefont {I.}~\bibnamefont
  {Ghebregziabher}}, \bibinfo {author} {\bibfnamefont {G.}~\bibnamefont
  {Golovin}}, \bibinfo {author} {\bibfnamefont {C.}~\bibnamefont {Liu}},
  \bibinfo {author} {\bibfnamefont {S.}~\bibnamefont {Chen}}, \bibinfo {author}
  {\bibfnamefont {S.}~\bibnamefont {Banerjee}}, \bibinfo {author}
  {\bibfnamefont {J.}~\bibnamefont {Zhang}}, \ and\ \bibinfo {author}
  {\bibfnamefont {D.~P.}\ \bibnamefont {Umstadter}},\ }\href@noop {} {\bibfield
   {journal} {\bibinfo  {journal} {Nature Photon.}\ }\textbf {\bibinfo {volume}
  {8}},\ \bibinfo {pages} {28} (\bibinfo {year} {2014})}\BibitemShut {NoStop}%
\bibitem [{\citenamefont {Albert}\ \emph {et~al.}(2010)\citenamefont {Albert},
  \citenamefont {Anderson}, \citenamefont {Anderson}, \citenamefont {Betts},
  \citenamefont {Gibson}, \citenamefont {Hagmann}, \citenamefont {Hall},
  \citenamefont {Johnson}, \citenamefont {Messerly}, \citenamefont {Semenov},
  \citenamefont {Shverdin}, \citenamefont {Tremaine}, \citenamefont
  {Hartemann}, \citenamefont {Siders}, \citenamefont {McNabb},\ and\
  \citenamefont {Barty}}]{Albert_2010l}%
  \BibitemOpen
  \bibfield  {author} {\bibinfo {author} {\bibfnamefont {F.}~\bibnamefont
  {Albert}}, \bibinfo {author} {\bibfnamefont {S.~G.}\ \bibnamefont
  {Anderson}}, \bibinfo {author} {\bibfnamefont {G.~A.}\ \bibnamefont
  {Anderson}}, \bibinfo {author} {\bibfnamefont {S.~M.}\ \bibnamefont {Betts}},
  \bibinfo {author} {\bibfnamefont {D.~J.}\ \bibnamefont {Gibson}}, \bibinfo
  {author} {\bibfnamefont {C.~A.}\ \bibnamefont {Hagmann}}, \bibinfo {author}
  {\bibfnamefont {J.}~\bibnamefont {Hall}}, \bibinfo {author} {\bibfnamefont
  {M.~S.}\ \bibnamefont {Johnson}}, \bibinfo {author} {\bibfnamefont {M.~J.}\
  \bibnamefont {Messerly}}, \bibinfo {author} {\bibfnamefont {V.~A.}\
  \bibnamefont {Semenov}}, \bibinfo {author} {\bibfnamefont {M.~Y.}\
  \bibnamefont {Shverdin}}, \bibinfo {author} {\bibfnamefont {A.~M.}\
  \bibnamefont {Tremaine}}, \bibinfo {author} {\bibfnamefont {F.~V.}\
  \bibnamefont {Hartemann}}, \bibinfo {author} {\bibfnamefont {C.~W.}\
  \bibnamefont {Siders}}, \bibinfo {author} {\bibfnamefont {D.~P.}\
  \bibnamefont {McNabb}}, \ and\ \bibinfo {author} {\bibfnamefont {C.~P.~J.}\
  \bibnamefont {Barty}},\ }\href@noop {} {\bibfield  {journal} {\bibinfo
  {journal} {Opt. Lett.}\ }\textbf {\bibinfo {volume} {35}},\ \bibinfo {pages}
  {354} (\bibinfo {year} {2010})}\BibitemShut {NoStop}%
\bibitem [{\citenamefont {Sarri}\ \emph {et~al.}(2014)\citenamefont {Sarri},
  \citenamefont {Corvan}, \citenamefont {Schumaker}, \citenamefont {Cole},
  \citenamefont {Di~Piazza}, \citenamefont {Ahmed}, \citenamefont {Harvey},
  \citenamefont {Keitel}, \citenamefont {Krushelnick}, \citenamefont {Mangles},
  \citenamefont {Najmudin}, \citenamefont {Symes}, \citenamefont {Thomas},
  \citenamefont {Yeung}, \citenamefont {Zhao},\ and\ \citenamefont
  {Zepf}}]{Sarri_2014}%
  \BibitemOpen
  \bibfield  {author} {\bibinfo {author} {\bibfnamefont {G.}~\bibnamefont
  {Sarri}}, \bibinfo {author} {\bibfnamefont {D.~J.}\ \bibnamefont {Corvan}},
  \bibinfo {author} {\bibfnamefont {W.}~\bibnamefont {Schumaker}}, \bibinfo
  {author} {\bibfnamefont {J.~M.}\ \bibnamefont {Cole}}, \bibinfo {author}
  {\bibfnamefont {A.}~\bibnamefont {Di~Piazza}}, \bibinfo {author}
  {\bibfnamefont {H.}~\bibnamefont {Ahmed}}, \bibinfo {author} {\bibfnamefont
  {C.}~\bibnamefont {Harvey}}, \bibinfo {author} {\bibfnamefont {C.~H.}\
  \bibnamefont {Keitel}}, \bibinfo {author} {\bibfnamefont {K.}~\bibnamefont
  {Krushelnick}}, \bibinfo {author} {\bibfnamefont {S.~P.~D.}\ \bibnamefont
  {Mangles}}, \bibinfo {author} {\bibfnamefont {Z.}~\bibnamefont {Najmudin}},
  \bibinfo {author} {\bibfnamefont {D.}~\bibnamefont {Symes}}, \bibinfo
  {author} {\bibfnamefont {A.~G.~R.}\ \bibnamefont {Thomas}}, \bibinfo {author}
  {\bibfnamefont {M.}~\bibnamefont {Yeung}}, \bibinfo {author} {\bibfnamefont
  {Z.}~\bibnamefont {Zhao}}, \ and\ \bibinfo {author} {\bibfnamefont
  {M.}~\bibnamefont {Zepf}},\ }\href@noop {} {\bibfield  {journal} {\bibinfo
  {journal} {Phys. Rev. Lett.}\ }\textbf {\bibinfo {volume} {113}},\ \bibinfo
  {pages} {224801} (\bibinfo {year} {2014})}\BibitemShut {NoStop}%
\bibitem [{\citenamefont {Goldman}(1964)}]{Goldman_1964}%
  \BibitemOpen
  \bibfield  {author} {\bibinfo {author} {\bibfnamefont {I.~I.}\ \bibnamefont
  {Goldman}},\ }\href@noop {} {\bibfield  {journal} {\bibinfo  {journal} {Zh.
  Eksp. Teor. Fiz.}\ }\textbf {\bibinfo {volume} {46}},\ \bibinfo {pages}
  {1412} (\bibinfo {year} {1964})},\ \bibinfo {note} {{[Sov. Phys. JETP 19, 954
  (1964)]}}\BibitemShut {NoStop}%
\bibitem [{\citenamefont {Brown}\ and\ \citenamefont
  {Kibble}(1964)}]{Brown_1964}%
  \BibitemOpen
  \bibfield  {author} {\bibinfo {author} {\bibfnamefont {L.~S.}\ \bibnamefont
  {Brown}}\ and\ \bibinfo {author} {\bibfnamefont {T.~W.}\ \bibnamefont
  {Kibble}},\ }\href@noop {} {\bibfield  {journal} {\bibinfo  {journal} {Phys.
  Rev.}\ }\textbf {\bibinfo {volume} {133}},\ \bibinfo {pages} {705} (\bibinfo
  {year} {1964})}\BibitemShut {NoStop}%
\bibitem [{\citenamefont {{Di Piazza}}\ \emph {et~al.}(2012)\citenamefont {{Di
  Piazza}}, \citenamefont {M\"uller}, \citenamefont {Hatsagortsyan},\ and\
  \citenamefont {Keitel}}]{RMP_2012}%
  \BibitemOpen
  \bibfield  {author} {\bibinfo {author} {\bibfnamefont {A.}~\bibnamefont {{Di
  Piazza}}}, \bibinfo {author} {\bibfnamefont {C.}~\bibnamefont {M\"uller}},
  \bibinfo {author} {\bibfnamefont {K.~Z.}\ \bibnamefont {Hatsagortsyan}}, \
  and\ \bibinfo {author} {\bibfnamefont {C.~H.}\ \bibnamefont {Keitel}},\
  }\href@noop {} {\bibfield  {journal} {\bibinfo  {journal} {Rev. Mod. Phys.}\
  }\textbf {\bibinfo {volume} {84}},\ \bibinfo {pages} {1177} (\bibinfo {year}
  {2012})}\BibitemShut {NoStop}%
\bibitem [{\citenamefont {{J. Koga and T. Zh. Esirkepov and S. V.
  Bulanov}}(2005)}]{Koga_2005}%
  \BibitemOpen
  \bibfield  {author} {\bibinfo {author} {\bibnamefont {{J. Koga and T. Zh.
  Esirkepov and S. V. Bulanov}}},\ }\href@noop {} {\bibfield  {journal}
  {\bibinfo  {journal} {Phys. Plasmas}\ }\textbf {\bibinfo {volume} {12}},\
  \bibinfo {pages} {093106} (\bibinfo {year} {2005})}\BibitemShut {NoStop}%
\bibitem [{\citenamefont {{Di~Piazza}}\ \emph {et~al.}(2009)\citenamefont
  {{Di~Piazza}}, \citenamefont {Hatsagortsyan},\ and\ \citenamefont
  {Keitel}}]{DiPiazza_2009}%
  \BibitemOpen
  \bibfield  {author} {\bibinfo {author} {\bibfnamefont {A.}~\bibnamefont
  {{Di~Piazza}}}, \bibinfo {author} {\bibfnamefont {K.~Z.}\ \bibnamefont
  {Hatsagortsyan}}, \ and\ \bibinfo {author} {\bibfnamefont {C.~H.}\
  \bibnamefont {Keitel}},\ }\href@noop {} {\bibfield  {journal} {\bibinfo
  {journal} {Phys. Rev. Lett.}\ }\textbf {\bibinfo {volume} {102}},\ \bibinfo
  {pages} {254802} (\bibinfo {year} {2009})}\BibitemShut {NoStop}%
\bibitem [{\citenamefont {{Di~Piazza}}\ \emph {et~al.}(2010)\citenamefont
  {{Di~Piazza}}, \citenamefont {Hatsagortsyan},\ and\ \citenamefont
  {Keitel}}]{DiPiazza_2010}%
  \BibitemOpen
  \bibfield  {author} {\bibinfo {author} {\bibfnamefont {A.}~\bibnamefont
  {{Di~Piazza}}}, \bibinfo {author} {\bibfnamefont {K.~Z.}\ \bibnamefont
  {Hatsagortsyan}}, \ and\ \bibinfo {author} {\bibfnamefont {C.~H.}\
  \bibnamefont {Keitel}},\ }\href@noop {} {\bibfield  {journal} {\bibinfo
  {journal} {Phys. Rev. Lett.}\ }\textbf {\bibinfo {volume} {105}},\ \bibinfo
  {pages} {220403} (\bibinfo {year} {2010})}\BibitemShut {NoStop}%
\bibitem [{\citenamefont {Sokolov}\ \emph
  {et~al.}(2010{\natexlab{a}})\citenamefont {Sokolov}, \citenamefont {Nees},
  \citenamefont {Yanovsky}, \citenamefont {Naumova},\ and\ \citenamefont
  {Mourou}}]{Sokolov_2010}%
  \BibitemOpen
  \bibfield  {author} {\bibinfo {author} {\bibfnamefont {I.~V.}\ \bibnamefont
  {Sokolov}}, \bibinfo {author} {\bibfnamefont {J.~A.}\ \bibnamefont {Nees}},
  \bibinfo {author} {\bibfnamefont {V.~P.}\ \bibnamefont {Yanovsky}}, \bibinfo
  {author} {\bibfnamefont {N.~M.}\ \bibnamefont {Naumova}}, \ and\ \bibinfo
  {author} {\bibfnamefont {G.~A.}\ \bibnamefont {Mourou}},\ }\href@noop {}
  {\bibfield  {journal} {\bibinfo  {journal} {Phys. Rev. E}\ }\textbf {\bibinfo
  {volume} {81}},\ \bibinfo {pages} {036412} (\bibinfo {year}
  {2010}{\natexlab{a}})}\BibitemShut {NoStop}%
\bibitem [{\citenamefont {Sokolov}\ \emph
  {et~al.}(2010{\natexlab{b}})\citenamefont {Sokolov}, \citenamefont {Naumova},
  \citenamefont {Nees},\ and\ \citenamefont {Mourou}}]{Sokolov_2010b}%
  \BibitemOpen
  \bibfield  {author} {\bibinfo {author} {\bibfnamefont {I.~V.}\ \bibnamefont
  {Sokolov}}, \bibinfo {author} {\bibfnamefont {N.~M.}\ \bibnamefont
  {Naumova}}, \bibinfo {author} {\bibfnamefont {J.~A.}\ \bibnamefont {Nees}}, \
  and\ \bibinfo {author} {\bibfnamefont {G.~A.}\ \bibnamefont {Mourou}},\
  }\href@noop {} {\bibfield  {journal} {\bibinfo  {journal} {Phys. Rev. Lett.}\
  }\textbf {\bibinfo {volume} {105}},\ \bibinfo {pages} {195005} (\bibinfo
  {year} {2010}{\natexlab{b}})}\BibitemShut {NoStop}%
\bibitem [{\citenamefont {Thomas}\ \emph {et~al.}(2012)\citenamefont {Thomas},
  \citenamefont {Ridgers}, \citenamefont {Bulanov}, \citenamefont {Griffin},\
  and\ \citenamefont {Mangles}}]{Thomas_2012}%
  \BibitemOpen
  \bibfield  {author} {\bibinfo {author} {\bibfnamefont {A.~G.~R.}\
  \bibnamefont {Thomas}}, \bibinfo {author} {\bibfnamefont {C.~P.}\
  \bibnamefont {Ridgers}}, \bibinfo {author} {\bibfnamefont {S.~S.}\
  \bibnamefont {Bulanov}}, \bibinfo {author} {\bibfnamefont {B.~J.}\
  \bibnamefont {Griffin}}, \ and\ \bibinfo {author} {\bibfnamefont {S.~P.~D.}\
  \bibnamefont {Mangles}},\ }\href@noop {} {\bibfield  {journal} {\bibinfo
  {journal} {Phys. Rev. X}\ }\textbf {\bibinfo {volume} {2}},\ \bibinfo {pages}
  {041004} (\bibinfo {year} {2012})}\BibitemShut {NoStop}%
\bibitem [{\citenamefont {Green}\ and\ \citenamefont
  {Harvey}(2014)}]{Green_2014}%
  \BibitemOpen
  \bibfield  {author} {\bibinfo {author} {\bibfnamefont {D.~G.}\ \bibnamefont
  {Green}}\ and\ \bibinfo {author} {\bibfnamefont {C.~N.}\ \bibnamefont
  {Harvey}},\ }\href@noop {} {\bibfield  {journal} {\bibinfo  {journal} {Phys.
  Rev. Lett.}\ }\textbf {\bibinfo {volume} {112}},\ \bibinfo {pages} {164801}
  (\bibinfo {year} {2014})}\BibitemShut {NoStop}%
\bibitem [{\citenamefont {Blackburn}\ \emph {et~al.}(2014)\citenamefont
  {Blackburn}, \citenamefont {Ridgers}, \citenamefont {Kirk},\ and\
  \citenamefont {Bell}}]{Blackburn_2014}%
  \BibitemOpen
  \bibfield  {author} {\bibinfo {author} {\bibfnamefont {T.~G.}\ \bibnamefont
  {Blackburn}}, \bibinfo {author} {\bibfnamefont {C.~P.}\ \bibnamefont
  {Ridgers}}, \bibinfo {author} {\bibfnamefont {J.~G.}\ \bibnamefont {Kirk}}, \
  and\ \bibinfo {author} {\bibfnamefont {A.~R.}\ \bibnamefont {Bell}},\
  }\href@noop {} {\bibfield  {journal} {\bibinfo  {journal} {Phys. Rev. Lett.}\
  }\textbf {\bibinfo {volume} {112}},\ \bibinfo {pages} {015001} (\bibinfo
  {year} {2014})}\BibitemShut {NoStop}%
\bibitem [{\citenamefont {Li}\ \emph {et~al.}(2014)\citenamefont {Li},
  \citenamefont {Hatsagortsyan},\ and\ \citenamefont
  {Keitel}}]{Jian-Xing_2014}%
  \BibitemOpen
  \bibfield  {author} {\bibinfo {author} {\bibfnamefont {J.-X.}\ \bibnamefont
  {Li}}, \bibinfo {author} {\bibfnamefont {K.~Z.}\ \bibnamefont
  {Hatsagortsyan}}, \ and\ \bibinfo {author} {\bibfnamefont {C.~H.}\
  \bibnamefont {Keitel}},\ }\href@noop {} {\bibfield  {journal} {\bibinfo
  {journal} {Phys. Rev. Lett.}\ }\textbf {\bibinfo {volume} {113}},\ \bibinfo
  {pages} {044801} (\bibinfo {year} {2014})}\BibitemShut {NoStop}%
\bibitem [{\citenamefont {{The Extreme Light Infrastructure (ELI)}}()}]{ELI}%
  \BibitemOpen
  \bibfield  {author} {\bibinfo {author} {\bibnamefont {{The Extreme Light
  Infrastructure (ELI)}}},\ }\href@noop {} {}\bibinfo {howpublished}
  {\url{http://www.eli-laser.eu/}}\BibitemShut {NoStop}%
\bibitem [{\citenamefont {{The High Power laser Energy Research (HiPER)
  facility}}()}]{HiPER}%
  \BibitemOpen
  \bibfield  {author} {\bibinfo {author} {\bibnamefont {{The High Power laser
  Energy Research (HiPER) facility}}},\ }\href {http://www.hiper-laser.org/}
  {}\BibitemShut {NoStop}%
\bibitem [{\citenamefont {Krausz}\ and\ \citenamefont
  {Ivanov}(2009)}]{Krausz_2009}%
  \BibitemOpen
  \bibfield  {author} {\bibinfo {author} {\bibfnamefont {F.}~\bibnamefont
  {Krausz}}\ and\ \bibinfo {author} {\bibfnamefont {M.}~\bibnamefont
  {Ivanov}},\ }\href@noop {} {\bibfield  {journal} {\bibinfo  {journal} {Rev.
  Mod. Phys.}\ }\textbf {\bibinfo {volume} {81}},\ \bibinfo {pages} {163}
  (\bibinfo {year} {2009})}\BibitemShut {NoStop}%
\bibitem [{\citenamefont {Ledingham}\ \emph {et~al.}(2003)\citenamefont
  {Ledingham}, \citenamefont {McKenna},\ and\ \citenamefont
  {Singhal}}]{Ledingham_2003}%
  \BibitemOpen
  \bibfield  {author} {\bibinfo {author} {\bibfnamefont {K.~W.~D.}\
  \bibnamefont {Ledingham}}, \bibinfo {author} {\bibfnamefont {P.}~\bibnamefont
  {McKenna}}, \ and\ \bibinfo {author} {\bibfnamefont {R.~P.}\ \bibnamefont
  {Singhal}},\ }\href@noop {} {\bibfield  {journal} {\bibinfo  {journal}
  {Science}\ }\textbf {\bibinfo {volume} {300}},\ \bibinfo {pages} {1107}
  (\bibinfo {year} {2003})}\BibitemShut {NoStop}%
\bibitem [{\citenamefont {P\'alffy}\ \emph {et~al.}()\citenamefont {P\'alffy},
  \citenamefont {Buss}, \citenamefont {Hoefer},\ and\ \citenamefont
  {Weidenm\"uller}}]{Palffy_2015}%
  \BibitemOpen
  \bibfield  {author} {\bibinfo {author} {\bibfnamefont {A.}~\bibnamefont
  {P\'alffy}}, \bibinfo {author} {\bibfnamefont {O.}~\bibnamefont {Buss}},
  \bibinfo {author} {\bibfnamefont {A.}~\bibnamefont {Hoefer}}, \ and\ \bibinfo
  {author} {\bibfnamefont {H.~A.}\ \bibnamefont {Weidenm\"uller}},\ }\href@noop
  {} {\ }\bibinfo {note} {{arXiv:1506.04127 [nucl-th] }}\BibitemShut {NoStop}%
\bibitem [{\citenamefont {Povh}\ \emph {et~al.}(1993)\citenamefont {Povh},
  \citenamefont {Rith}, \citenamefont {Scholz},\ and\ \citenamefont
  {Zetsche}}]{Povh_book}%
  \BibitemOpen
  \bibfield  {author} {\bibinfo {author} {\bibfnamefont {B.}~\bibnamefont
  {Povh}}, \bibinfo {author} {\bibfnamefont {K.}~\bibnamefont {Rith}}, \bibinfo
  {author} {\bibfnamefont {C.}~\bibnamefont {Scholz}}, \ and\ \bibinfo {author}
  {\bibfnamefont {F.}~\bibnamefont {Zetsche}},\ }\href@noop {} {\emph {\bibinfo
  {title} {Particle and Nuclei}}}\ (\bibinfo  {publisher} {Springer, Berlin},\
  \bibinfo {year} {1993})\BibitemShut {NoStop}%
\bibitem [{Sup()}]{Suppl_material}%
  \BibitemOpen
  \href@noop {} {}\bibinfo {howpublished} {see the details in the Supplemental
  Material.}\BibitemShut {Stop}%
\bibitem [{\citenamefont {Salamin}\ and\ \citenamefont
  {Faisal}(1996)}]{Salamin_1996}%
  \BibitemOpen
  \bibfield  {author} {\bibinfo {author} {\bibfnamefont {Y.~I.}\ \bibnamefont
  {Salamin}}\ and\ \bibinfo {author} {\bibfnamefont {F.~H.~M.}\ \bibnamefont
  {Faisal}},\ }\href@noop {} {\bibfield  {journal} {\bibinfo  {journal} {Phys.
  Rev. A}\ }\textbf {\bibinfo {volume} {54}},\ \bibinfo {pages} {4383}
  (\bibinfo {year} {1996})}\BibitemShut {NoStop}%
\bibitem [{Bau()}]{Baurichter_1997_new}%
  \BibitemOpen
  \href@noop {} {\ }\bibinfo {note} {Note that the effect of radiation reaction
  for multi-GeV electrons in oriented crystals has been well investigated in
  the classical and quantum regimes, respectively, see, e.g., A. Baurichter, K.
  Kirsebom, Yu. V. Kononets, R. Medenwaldt, U. Mikkelsen, S. P. M$\o$ller, E.
  Uggerh$\o$j, T. Worm, K. Elsener, S. Ballestrero, P. Sona, J. Romano, S. H.
  Connell, J. P. F. Sellschop, R. O. Avakian, A. E. Avetisian, and S. P.
  Taroianet, Phys. Rev. Lett. {\bf 79}, 3415 (1997).}\BibitemShut {Stop}%
\bibitem [{\citenamefont {Elkina}\ \emph {et~al.}(2011)\citenamefont {Elkina},
  \citenamefont {Fedotov}, \citenamefont {Kostyukov}, \citenamefont {Legkov},
  \citenamefont {Narozhny}, \citenamefont {Nerush},\ and\ \citenamefont
  {Ruhl}}]{Elkina_2011}%
  \BibitemOpen
  \bibfield  {author} {\bibinfo {author} {\bibfnamefont {N.~V.}\ \bibnamefont
  {Elkina}}, \bibinfo {author} {\bibfnamefont {A.~M.}\ \bibnamefont {Fedotov}},
  \bibinfo {author} {\bibfnamefont {I.~Y.}\ \bibnamefont {Kostyukov}}, \bibinfo
  {author} {\bibfnamefont {M.~V.}\ \bibnamefont {Legkov}}, \bibinfo {author}
  {\bibfnamefont {N.~B.}\ \bibnamefont {Narozhny}}, \bibinfo {author}
  {\bibfnamefont {E.~N.}\ \bibnamefont {Nerush}}, \ and\ \bibinfo {author}
  {\bibfnamefont {H.}~\bibnamefont {Ruhl}},\ }\href@noop {} {\bibfield
  {journal} {\bibinfo  {journal} {Phys. Rev. ST Accel. Beams}\ }\textbf
  {\bibinfo {volume} {14}},\ \bibinfo {pages} {054401} (\bibinfo {year}
  {2011})}\BibitemShut {NoStop}%
\bibitem [{\citenamefont {Ridgers}\ \emph {et~al.}(2014)\citenamefont
  {Ridgers}, \citenamefont {Kirk}, \citenamefont {Duclous}, \citenamefont
  {Blackburn}, \citenamefont {Brady}, \citenamefont {Bennett}, \citenamefont
  {Arber},\ and\ \citenamefont {Bell}}]{Ridgers_2014}%
  \BibitemOpen
  \bibfield  {author} {\bibinfo {author} {\bibfnamefont {C.~P.}\ \bibnamefont
  {Ridgers}}, \bibinfo {author} {\bibfnamefont {J.~G.}\ \bibnamefont {Kirk}},
  \bibinfo {author} {\bibfnamefont {R.}~\bibnamefont {Duclous}}, \bibinfo
  {author} {\bibfnamefont {T.~G.}\ \bibnamefont {Blackburn}}, \bibinfo {author}
  {\bibfnamefont {C.~S.}\ \bibnamefont {Brady}}, \bibinfo {author}
  {\bibfnamefont {K.}~\bibnamefont {Bennett}}, \bibinfo {author} {\bibfnamefont
  {T.~D.}\ \bibnamefont {Arber}}, \ and\ \bibinfo {author} {\bibfnamefont
  {A.~R.}\ \bibnamefont {Bell}},\ }\href@noop {} {\bibfield  {journal}
  {\bibinfo  {journal} {J. Compt. Phys.}\ }\textbf {\bibinfo {volume} {260}},\
  \bibinfo {pages} {273} (\bibinfo {year} {2014})}\BibitemShut {NoStop}%
\bibitem [{\citenamefont {Green}\ and\ \citenamefont
  {Harvey}(2015)}]{Green_2015}%
  \BibitemOpen
  \bibfield  {author} {\bibinfo {author} {\bibfnamefont {D.~G.}\ \bibnamefont
  {Green}}\ and\ \bibinfo {author} {\bibfnamefont {C.~N.}\ \bibnamefont
  {Harvey}},\ }\href@noop {} {\bibfield  {journal} {\bibinfo  {journal} {Comp.
  Phys. Commun.}\ }\textbf {\bibinfo {volume} {192}},\ \bibinfo {pages} {313}
  (\bibinfo {year} {2015})}\BibitemShut {NoStop}%
\bibitem [{\citenamefont {Ritus}(1985)}]{Ritus_1985}%
  \BibitemOpen
  \bibfield  {author} {\bibinfo {author} {\bibfnamefont {V.~I.}\ \bibnamefont
  {Ritus}},\ }\href@noop {} {\bibfield  {journal} {\bibinfo  {journal} {J. Sov.
  Laser Res.}\ }\textbf {\bibinfo {volume} {6}},\ \bibinfo {pages} {497}
  (\bibinfo {year} {1985})}\BibitemShut {NoStop}%
\bibitem [{\citenamefont {Khokonov}\ and\ \citenamefont
  {Bekulova}(2010)}]{Khokonov_2010}%
  \BibitemOpen
  \bibfield  {author} {\bibinfo {author} {\bibfnamefont {M.~K.}\ \bibnamefont
  {Khokonov}}\ and\ \bibinfo {author} {\bibfnamefont {I.}~\bibnamefont
  {Bekulova}},\ }\href@noop {} {\bibfield  {journal} {\bibinfo  {journal}
  {Techn. Phys.}\ }\textbf {\bibinfo {volume} {55}},\ \bibinfo {pages} {728}
  (\bibinfo {year} {2010})}\BibitemShut {NoStop}%
\bibitem [{\citenamefont {Baier}\ \emph {et~al.}(1994)\citenamefont {Baier},
  \citenamefont {Katkov},\ and\ \citenamefont {Strakhovenko}}]{Baier_b_1994}%
  \BibitemOpen
  \bibfield  {author} {\bibinfo {author} {\bibfnamefont {V.~N.}\ \bibnamefont
  {Baier}}, \bibinfo {author} {\bibfnamefont {V.~M.}\ \bibnamefont {Katkov}}, \
  and\ \bibinfo {author} {\bibfnamefont {V.~M.}\ \bibnamefont {Strakhovenko}},\
  }\href@noop {} {\emph {\bibinfo {title} {Electromagnetic Processes at High
  Energies in Oriented Single Crystals}}}\ (\bibinfo  {publisher} {World
  Scientific, Singapore},\ \bibinfo {year} {1994})\BibitemShut {NoStop}%
\bibitem [{\citenamefont {Khokonov}\ and\ \citenamefont
  {Nitta}(2002)}]{Khokonov_2002a}%
  \BibitemOpen
  \bibfield  {author} {\bibinfo {author} {\bibfnamefont {M.~K.}\ \bibnamefont
  {Khokonov}}\ and\ \bibinfo {author} {\bibfnamefont {H.}~\bibnamefont
  {Nitta}},\ }\href@noop {} {\bibfield  {journal} {\bibinfo  {journal} {Phys.
  Rev. Lett.}\ }\textbf {\bibinfo {volume} {89}},\ \bibinfo {pages} {094801}
  (\bibinfo {year} {2002})}\BibitemShut {NoStop}%
\end{thebibliography}%

\end{document}